\documentclass[prd,twocolumn,nofootinbib]{revtex4}
\usepackage{epsfig,psfrag,graphics,color,verbatim}
\usepackage{amsmath}
\def\s#1{\widetilde{#1}}
\def\neut{{\tilde\chi_{1}^0}}
\def\relic{\Omega_{\neut}}
\def\crosssection{\sigma_{\chi-\mbox{\tiny{p}}}^{\mbox{\tiny{SI}}}}

\begin{document}

\title{Identification of Dark Matter particles with LHC and direct detection data} 
\author{G.~Bertone\,$^{1,2}$ D.G.~Cerde\~no\,$^{3}$ M.~Fornasa\,$^{4}$ R.~Ruiz de Austri\,$^{5}$ and  R.~Trotta\,$^{6}$}

\affiliation{${^1}$ Institut for Theoretical Physics, Univ. of Z\"urich, Winterthurerst. 190, 8057 Z\"urich CH} 
\affiliation{${^2}$ Institut d'Astrophysique de Paris, UMR 7095-CNRS, Univ. P. et M. Curie, 98bis Bd Arago, 75014 Paris, France} 
\affiliation{${^3}$ Departamento de F\'{\i}sica Te\'{o}rica, and Instituto de F\'{\i}sica Te\'{o}rica UAM/CSIC, Universidad Aut\'{o}noma de Madrid, Cantoblanco, E-28049 Madrid, Spain}
\affiliation{${^4}$ Instituto de Astrof\'{\i}sica de Andaluc\'{\i}a (CSIC), E-18008, Granada, Spain} 
\affiliation{${^5}$ Instituto de F\'isica Corpuscular, IFIC-UV/CSIC, Valencia, Spain} 
\affiliation{${^6}$ Astrophysics Group, Imperial College London, Blackett Laboratory, Prince Consort Road, London SW7 2AZ, UK} 

\begin{abstract}
Dark matter (DM) is currently searched for with a variety of detection 
strategies. Accelerator searches are particularly promising, but even if 
Weakly Interacting Massive Particles (WIMPs) are found at the Large Hadron 
Collider (LHC), it will be difficult to prove that they constitute the bulk 
of the DM in the Universe $ \Omega_{\rm \tiny{DM}} $. We show that a significantly better reconstruction of the DM properties can be obtained with a combined analysis of LHC and direct detection (DD) data, by making a simple Ansatz on the WIMP local density $ \rho_{\neut} $, i.e., by  
assuming that the local densiy scales with the cosmological relic abundance, $ (\rho_{\neut}/\rho_{\rm \tiny{DM}})=(\relic/\Omega_{\rm \tiny{DM}} $). 
 We demonstrate this method in 
an explicit example in the context of a 24-parameter supersymmetric model, 
with a neutralino LSP in the stau co-annihilation region. Our results show 
that future ton-scale DD experiments will allow to break 
degeneracies in the SUSY parameter space and achieve a significantly better reconstruction 
of the neutralino composition and its relic density than with LHC data alone.
\end{abstract}

\maketitle

\section{Introduction}
\label{sec:introduction}
Identifying the nature of the dark matter (DM) remains one of the central 
unsolved problems in modern particle physics and cosmology. A generic 
Weakly Interacting Massive Particle (WIMP) is among the best-motivated 
possibilities since it can be thermally produced in the early Universe in 
the right amount to account for the observed DM density. Indeed, many theories 
for Physics beyond the Standard Model contain viable WIMP candidates, as is 
the case of Supersymmetry (SUSY) when the lightest SUSY particle (LSP) is 
the lightest neutralino  (a linear superposition of the supersymmetric
partners of the gauge and Higgs bosons) 
\cite{Jungman:1995df,Munoz:2003gx,Bertone:2004pz,book}.

DM can be searched for in various ways. One possibility is attempting a 
{\it direct} detection, through its scattering off nuclei inside an 
underground detector. Many experiments have been running or are under 
construction which are mostly sensitive to the spin-independent part of the 
WIMP-nucleus cross section, $ \crosssection $. Among these, the DAMA/LIBRA 
collaboration reported a possible DM signal 
\cite{Bernabei:2008yi,Bottino:2008mf}. However, its interpretation in terms 
of the elastic scattering of a WIMP with a mass around $ 10-100 \mbox{ GeV} $ and 
$ \crosssection \sim 10^{-3}-10^{-5}\, \mbox{pb} $ has been challenged by 
other experiments, such as the CoGeNT \cite{Aalseth:2008rx,Aalseth:2010vx}, 
CDMS \cite{Ahmed:2008eu,Ahmed:2009zw}, XENON \cite{Angle:2007uj} and ZEPLIN \cite{Lebedenko:2009xe}.
The CoGeNT collaboration has itself recently reported an irreducible excess of low-energy events 
which could also be understood as due 
to the scattering of a very light WIMP \cite{Aalseth:2010vx} (see also 
Ref.\,\cite{Chang:2010yk}), but this intepretation has in turn be put under pressure by the 
XENON-100 results, obtained with a fiducial target mass of 40 kg and 11 
days of exposure~\cite{Aprile:2010um}. 
Finally, the recent results from the CDMS-II collaboration 
show two events compatible with a WIMP signal, although these results are 
still statistically inconclusive \cite{Ahmed:2009zw}. 

The future increase of the sensitivity 
may clarify the situation, but it is becoming clear that several independent pieces of evidence 
will be necessary to claim discovery of DM. In fact, even if in principle the WIMP mass and scattering cross
section can be determined with some accuracy after its direct detection in 
one direct detection experiment, provided that the measured event rate is large and the WIMP mass is small \cite{Green:2007rb,Green:2008rd},
a second direct detection 
with a different target would actually allow a much more precise determination of the 
WIMP mass \cite{Drees:2008bv}, and if the new target is sensitive to the 
spin-dependent contribution of the WIMP-nucleus cross section it could even 
be used to discriminate among WIMP candidates \cite{Bertone:2007xj}. 

Another possibility consists in looking for the products of DM annihilation 
(e.g., high energy neutrinos, gamma-rays or antimatter) and thus 
{\it indirectly} reveal the presence of the DM \cite{Bertone:2004pz,book}. We leave the discussion of 
this search strategy to a forthcoming work, where we will present the 
constraints that can be set on the DM parameter space from the observation 
(or non-observation, see also \cite{Scott:2009jn}) of DM annihilation 
radiation \cite{Bertone:2010}.

Finally, collider experiments, most notably the Large Hadron Collider (LHC), 
will explore the nature of Physics at the TeV scale, where many of the extensions of the SM that propose DM 
candidates would manifest themselves. 
The detection of new Physics in particle colliders can provide crucial 
information about DM. For example, the mass and spin of the LSP could be 
determined through the study of kinematic variables 
\cite{Cho:2007qv,Cho:2008tj}. However, to prove that the newly discovered
particles account for all (or most) of the DM in the Universe, is a 
challenging task. In fact, although particle accelerators can provide some 
information about the neutralino relic density \cite{Baltz:2006fm}, it was 
found that 
in many cases the LHC would be unable to determine the precise composition 
of the neutralino, leading to an unreliable prediction of its relic 
abundance or to the occurrence of multiple solutions spanning several orders 
of magnitude, thus not allowing to establish whether or not it is the DM (see also Ref. \cite{Nath:2010zj} and references therein). 

One possibility is to build a new collider, such as the proposed 
International Linear Collider (ILC), that would allow a much more precise 
evaluation of the supersymmetric masses and couplings, and a better 
determination of the inferred relic density, as argued by the authors of 
Ref.\,\cite{Baltz:2006fm}. However, this machine will not be available in 
the near future, and it is therefore crucial to devise strategies that can 
be implemented as soon as new particles are discovered at the LHC. 
Fortunately, direct detection experiments are expected to greatly improve 
their sensitivity in the next few years and start probing interesting regions 
of the supersymmetric parameter space. In case of discovery, it will certainly be reassuring
if the mass reconstructed from direct detection experiments matched the value obtained from
accelerator measurements, since it would prove the existence of a particle which is stable over
cosmological timescales. The error on the mass reconstructed from direct detection experiments depends on the DM particle parameters, and on the experimental setup, and the interested reader can find a detailed analysis in Refs.\cite{Green:2007rb,Green:2008rd}. But one can do much more than checking the compatibility of
the two mass determinations. We show here that a combined analysis of the two data sets will allow a 
much better reconstruction of the DM properties, and a convincing 
identification of DM particles.  

Although the strategy discussed here is model-independent, we work out an 
explicit example in the context of a 24-parameters supersymmetric model, 
with a neutralino LSP in the stau co-annihilation region.


\section{Theoretical framework and LHC data} 
We work within the framework of the minimal supersymmetric extension
of the Standard Model (MSSM), for which we adopt a low energy parametrization 
in terms of 24 parameters, corresponding to its CP-conserving version. 
The input parameters are the coefficients of the trilinear terms for the 
three generations, the mass terms for gauginos (for which no universality 
assumption is made), right-handed and left-handed squarks and leptons, the 
mass of the pseudoscalar Higgs, the Higgsino mass parameter $ \mu $, and 
finally the ratio between the vacuum expectation values of the two Higgs 
bosons $ \tan\beta $.

If searches for new Physics at the LHC are consistent with a SUSY scenario,
the study of different kinematical variables will allow us to determine some 
properties of the SUSY spectrum. In particular, the masses of several
particles or mass-splittings between them could be extracted, with a 
precision that obviously depends on the properties of the specific point of 
the parameter space. These measurements can then be used as constraints on the 
24-dimensional SUSY model, in order to determine the regions of the MSSM parameter space which 
are consistent with such a measurement. This can be done by applying Bayes' theorem
\begin{equation}
p(\mathbf{x}|\mathbf{d})=
\frac{p(\mathbf{d}|\mathbf{x})p(\mathbf{x})}{p(\mathbf{d})},
\label{eqn:Bayesian_theorem}
\end{equation}
which updates the so-called prior probability density $ p(\mathbf{x}) $, encapsulating the knowledge of the 
24-dimensional space before taking into account the experimental constraints, $ \mathbf{d} $, into the posterior probability function (pdf) 
$ p(\mathbf{x}|\mathbf{d}) $. The latter describes the probability density assigned to a generic 24-dimensional point 
$ \mathbf{x} $ once the data have been taken into account via the likelihood function 
$ p(\mathbf{d}|\mathbf{x})$. Furthermore, on the RHS of Eq.\,\eqref{eqn:Bayesian_theorem}, $ p(\mathbf{d}) $ is the Bayesian evidence which, in our case, can be
dropped since it simply plays the role of a normalization constant for the posterior in this context (see~\cite{Trotta:2008qt} for further details). 

The marginal pdf of a particular subset (as e.g. only one) of the 24 parameters 
defining $ \mathbf{x} $ can be obtained by integrating over the remaining 
directions:
\begin{equation}
p(x^i|\mathbf{d})=\int_{[1,24]\backslash \{i\}}p(\mathbf{x}|\mathbf{d}) 
dx^1... dx^{i-1}dx^{i+1}...dx^{24}.
\label{eqn:1dim_pdf}
\end{equation}

The posterior encodes both the information contained in the priors and in the
experimental constraints, but, ideally, it should be largely independent of the choice of priors, so that the posterior inference is dominated by the data contained in the likelihood. If some residual dependence on the prior
$ p(\mathcal{\mathbf{x}}) $ remains this should be considered as a sign that 
the experimental data employed are not constraining enough to override completely different plausible prior choices and therefore the resulting posterior should be interpreted with some care, as it might depend on the prior assumptions. The probability distribution for any observable that is a function
of the 24 SUSY parameters $ f(\mathbf{x})$ can also be obtained since
$ p(f|\mathbf{d})=\delta(f-f(\mathbf{x})) p(\mathbf{x}|\mathbf{d}) $.

For the practical implementation of the Bayesian analysis sketched above we employed the 
\texttt{SuperBayeS} code \cite{SuperBayeS},  extending the publicly available version 1.35 to handle the 24 dimensions of our SUSY 
parameter space. To scan in an efficient way the SUSY parameter space we have upgraded the MultiNest~\cite{Feroz:2008xx, MN} algorithm included in SuperBayeS to the latest MultiNest release (v 2.7). MultiNest is a multi-modal implementation of the nested sampling algorithm, which is used to produce a list of samples in parameter space whose density is proportional to the posterior pdf of Eq.~\eqref{eqn:Bayesian_theorem}. For further information on nested sampling we refer the reader to the appendix of
Ref.\,\cite{Trotta:2008bp} and references therein.

For the present work we have chosen a specific benchmark point in
the MSSM parameter space, corresponding to the low-energy extrapolation of 
model LCC3 defined in Ref.\,\cite{Baltz:2006fm}. This benchmark is 
representative of SUSY models in the co-annihilation region, where the 
lightest neutralino is almost degenerate in mass with the lightest stau. 
In this region, co-annihilation effects reduce the neutralino relic abundance down to 
values compatible with the results from the WMAP satellite 
\cite{Komatsu:2008hk}, and therefore, the mass difference between the 
neutralino and the lightest stau is a fundamental parameter for the 
reconstruction of the relic density.  
It has been shown  \cite{Baltz:2006fm} that for this benchmark point LHC would be able 
to provide a measurement of the masses of a good part of the SUSY 
spectrum, including the two lightest neutralinos (see Ref.~\cite{Khotilovich:2005gb} for an 
extension of this analysis to the case of the ILC).
However the masses of some particles (most notably the two heaviest 
neutralinos and both charginos) would not be measured. 
The set of measurements that we use as constraints in our analysis
corresponds to that in Table 6 of Ref.\,\cite{Baltz:2006fm} \footnote{The 
exact values that we are using are slightly different from those in Table 6 
of Ref.\,\cite{Baltz:2006fm} since we are not deriving the mass spectrum from 
the low energy extrapolation of a cMSSM point. On the contrary our reference 
SUSY model is defined at low energy to be near LCC3.}, which assumes an integrated luminosity of 300 fb$^{-1}$.
Furthermore, as pointed 
out in Ref.\,\cite{Arnowitt:2008bz}, the neutralino-stau mass difference can 
be measured with an accuracy of 20\% with 10 fb$^{-1}$ luminosity in models where the squark masses are 
much larger than those of the lightest chargino and second-lightest 
neutralino, as is our case. We therefore also include a measurement of the neutralino-stau mass difference in our likelihood. For convenience, we summarize in 
Table \ref{tab:constraints} the set of LHC measurements on which we build our likelihood. Each of the constraints listed in Table \ref{tab:constraints} is implemented in the likelihood as an independent Gaussian distributed measurement around the true value $\mu$ for that observable, with standard deviation $\sigma$, as given in Table~\ref{tab:constraints}.

\begin{table}
\centering
\begin{tabular}{lcc}
\hline
Mass & Benchmark value, $\mu$ \quad  & \quad LHC error, $\sigma$\quad  \\
\hline
$ m(\s\chi^0_1) $ & 139.3 & 14.0 \\
$ m(\s\chi^0_2) $ & 269.4 & 41.0  \\
$ m(\s e_1) $ & 257.3 & 50.0 \\
$ m(\s \mu_1) $ & 257.2 & 50.0 \\
$ m(h) $ & 118.50 & 0.25 \\
$ m(A) $ & 432.4 & 1.5 \\
$ m(\s \tau_1)- m(\s\chi^0_1) $ & 16.4 & 2.0 \\
$ m(\s u_R) $ & 859.4 & 78.0 \\
$ m(\s d_R) $ & 882.5 & 78.0 \\
$ m(\s s_R) $ & 882.5 & 78.0 \\
$ m(\s c_R) $ & 859.4 & 78.0 \\
$ m(\s u_L) $ & 876.6 & 121.0 \\
$ m(\s d_L) $ & 884.6 & 121.0 \\
$ m(\s s_L) $ & 884.6 & 121.0 \\
$ m(\s c_L) $ & 876.6 & 121.0 \\
$ m(\s b_1) $ & 745.1 & 35.0 \\
$ m(\s b_2) $ & 800.7 & 74.0 \\
$ m(\s t_1) $ & 624.9 & 315.0  \\
$ m(\s g) $ & 894.6 & 171.0 \\
$ m(\s e_2) $ & 328.9 & 50.0 \\
$ m(\s \mu_2) $ & 328.8 & 50.0 \\
\hline
\end{tabular}
\caption{Sparticle spectrum (in GeV) for our benchmark SUSY point and relative estimated measurements errors at the LHC (standard deviation $\sigma$).}  
\label{tab:constraints}
\end{table}

Compared with previous Bayesian studies in which only precision tests of the Standard Model are considered as experimental constraints  
\cite{deAustri:2006pe,Trotta:2008bp,AbdusSalam:2009qd,Cabrera:2009dm}, we are assuming here a scenario in
which LHC reports a quite stringent collection of measurements. For this reason 
our posterior constraints are quite tight and we expect the prior dependence of our results to be very mild. This is confirmed by the inspection of the profile likelihood, which agrees well with the posterior pdf (see \cite{Trotta:2008bp} for a detailed discussion). This indicates that volume effects from the prior are unlikely to be playing a major role given the strong constraints we assume for our benchmark point.

\section{Future direct detection data}
In the simulation of a direct detection experiment we assume a future 
signal giving a WIMP detection, namely a certain number of 
events $ N $ and a corresponding set of recoil energies 
$ \{ E_i \}_{i=1,...,N} $. The total number $ N $ of simulated events
is the  sum of both background events (mainly interactions of detector
nuclei with  neutrons from surrounding rock, from residual
contaminants or from spallation  of cosmic muons) and recoils due to
DM. 
For concreteness, we will exemplify the method in the case of an experiment akin to the 1-ton scale SuperCDMS
experiment \cite{CDMS}. We simulated the differential number of background events as
in  Ref.\,\cite{Bernal:2008zk}. Since the capability of a simulated direct detection experiment to 
reconstruct the DM properties (see 
Refs. \cite{Green:2007rb,Green:2008rd,Bernal:2008zk} for more details) is 
worse in the case of a constant background distribution than for an 
exponential one, we only consider 
the case of energy-independent background recoil spectrum in order to be conservative. Therefore, we adopt a constant background differential spectrum 
$ (dN_{\rm back}/dE) = {\rm const}$ which is normalized so that, when binning the spectrum
in 9 bins of 10 keV width (from $ E_{\rm th}=10 \mbox{ keV} $ to 
$ E_{\rm max}=100 \mbox{ keV}$) the number of background events in the first bin
is the same as the number of DM signal events there. 

The expected number of events $ \lambda $ for our benchmark model and 
for an exposure $ \epsilon=300 \mbox{ ton day} $ is obtained by integrating 
the sum of the differential rate of WIMP and background events
\begin{equation}
\lambda=\epsilon \int_{E_{\rm th}}^{E_{\rm max}}
\frac{dR_{\chi}}{dE}+\frac{dR_{\rm back}}{dE} \, dE. 
\label{eqn:number_of_events}
\end{equation}

The dependency of the WIMP event rate on the physical quantities in 
the problem becomes apparent in the following parametrization
\cite{Lewin:1995rx} 
\begin{equation} 
\frac{dR_{\chi}}{dE} = c_1R_0 e^{-E/(E_0c_2)} F^2(E)\, ,
\end{equation}
where 
\begin{equation} \label{eq:R0}
R_0=\frac{\crosssection \rho_\chi A^2 c^2 (m_\chi+m_p)^2}
{\sqrt{\pi} m^3 m_p^2 v_0}\,,
\end{equation}
and
\begin{equation}
E_0=\frac{2m_\chi^2v_0^2 Am_p}{(m_\chi+Am_p)^2c^2}\,.
\end{equation}
Here, $ \rho_\chi $ is the local WIMP density, $ A $ is the mass number  
of the target nuclei ($ A=73 $ in the case of Germanium), $ m_p $ is the  
proton mass, $ v_0 $ is the characteristic WIMP velocity and $ F^2(E) $  
denotes the nuclear form factor. A discussion on the values of the parameters 
$ c_1 $ and $ c_2 $ and the functional form of $ F(E) $ can be found in
Refs. \cite{Green:2007rb,Green:2008rd,Lewin:1995rx}. The specific 
values of the quantities for our case study are summarized in Table 
\ref{tab:DD_parameters}. 

\begin{table}
\centering
\begin{tabular}{c|c|c|c|c|c|c}
\hline
Target  & A & $ \epsilon $ & $ E_{\rm th} $ & $ E_{\rm max} $ & $ \rho_\chi $ & $ \lambda $ \\
\hline
Ge & 73 & $ 300\, \mbox{ton day} $ & 10 keV & 100 keV & 0.385 GeV cm$^{-3} $ & 638 \\
\hline
\end{tabular}
\caption{Relevant quantities for a SuperCDMS-like direct detection experiment. The quantity $\lambda$ gives the expected number of WIMP recoils for our SUSY benchmark model.}
\label{tab:DD_parameters}
\end{table}

In order to combine the result of a direct detection experiment with LHC data,
we run an additional scan of the SUSY parameter space 
including in the likelihood function an additional Poisson-distributed term that 
compares the number of events and their spectral shape predicted in each point in parameter space with the recoil spectrum corresponding to the 
benchmark value of Table \ref{tab:DD_parameters}. The overall background rate and its spectral shape are assumed to be known.

As shown by Eqs.\,\eqref{eqn:number_of_events}-\eqref{eq:R0}, the number of detected events 
is proportional to the product of the WIMP-proton cross section and the local
DM density $ \lambda \propto \crosssection \rho_\chi $. Therefore, unless one 
specifies the value of $ \rho_\chi $, any information on the number of events
leaves the scattering cross section practically unconstrained. 

We propose two different strategies to specify $ \rho_\chi $:
\begin{enumerate}
\item {\it Consistency check:} we impose that 
\begin{equation}
\rho_\chi=\rho_{\rm \tiny{DM}} \, ,
\end{equation}
and we adopt for this quantity the value obtained in a recent paper by
Catena and  Ullio \cite{Catena:2009mf}, through a careful analysis of
dynamical observables in the Galaxy, namely 
$ \rho_\chi=0.385 \mbox{ GeV} \mbox{ cm}^{-3} $ (see also \cite{Strigari:2009zb,Salucci:2010qr,Weber:2009pt,Pato:2010yq}). Although this assumption 
completely removes the degeneracy between $ \crosssection $ and $ \rho_\chi $, 
it forces the identification of neutralino with the DM particle, 
irrespectively of the value of its thermal relic density. 
This is therefore equivalent to assuming that, a non-standard cosmological 
history of the Universe can correct any excess or deficit in the thermal 
relic density and make it agree with the WMAP result, for example, either 
by invoking late injection of entropy, non-thermal production through 
late-decaying particles (such as a modulus or a gravitino 
\cite{Moroi:1999zb}), 
scenarios with a low-reheating temperature
\cite{Giudice:2000ex}
(see also Ref.\,\cite{Fornengo:2002db}) or a faster expansion rate
\cite{Salati:2002md,Profumo:2003hq}.   
For these reasons this Ansatz must be considered as a {\it consistency check} 
rather than a proof of the identification of DM particles.
\item {\it Scaling Ansatz:} we assume that the local density of the 
neutralino scales with the cosmological abundance. More precisely, we propose 
the following {\it Ansatz}
\begin{equation}
\rho_{\neut}/\rho_{\rm \tiny{DM}}=\relic/\Omega_{\rm \tiny{DM}}.
\end{equation}
This Ansatz is strictly valid in the reasonable case where the distribution of 
neutralinos in large structures, and in particular in the Galaxy, traces the 
cosmological distribution of the DM. This Ansatz is obviously true if neutralinos 
contribute all the DM in the Universe, but is also valid in the case where 
the neutralino is a subdominant component of DM, provided that DM behaves,
as expected, as a cold collisionless particle. As shown below, this
simple assumption is powerful tool to remove degeneracies in
the parameter space. 
\end{enumerate}
\begin{figure*}[t]
\hspace*{-0.5cm}
\epsfig{file=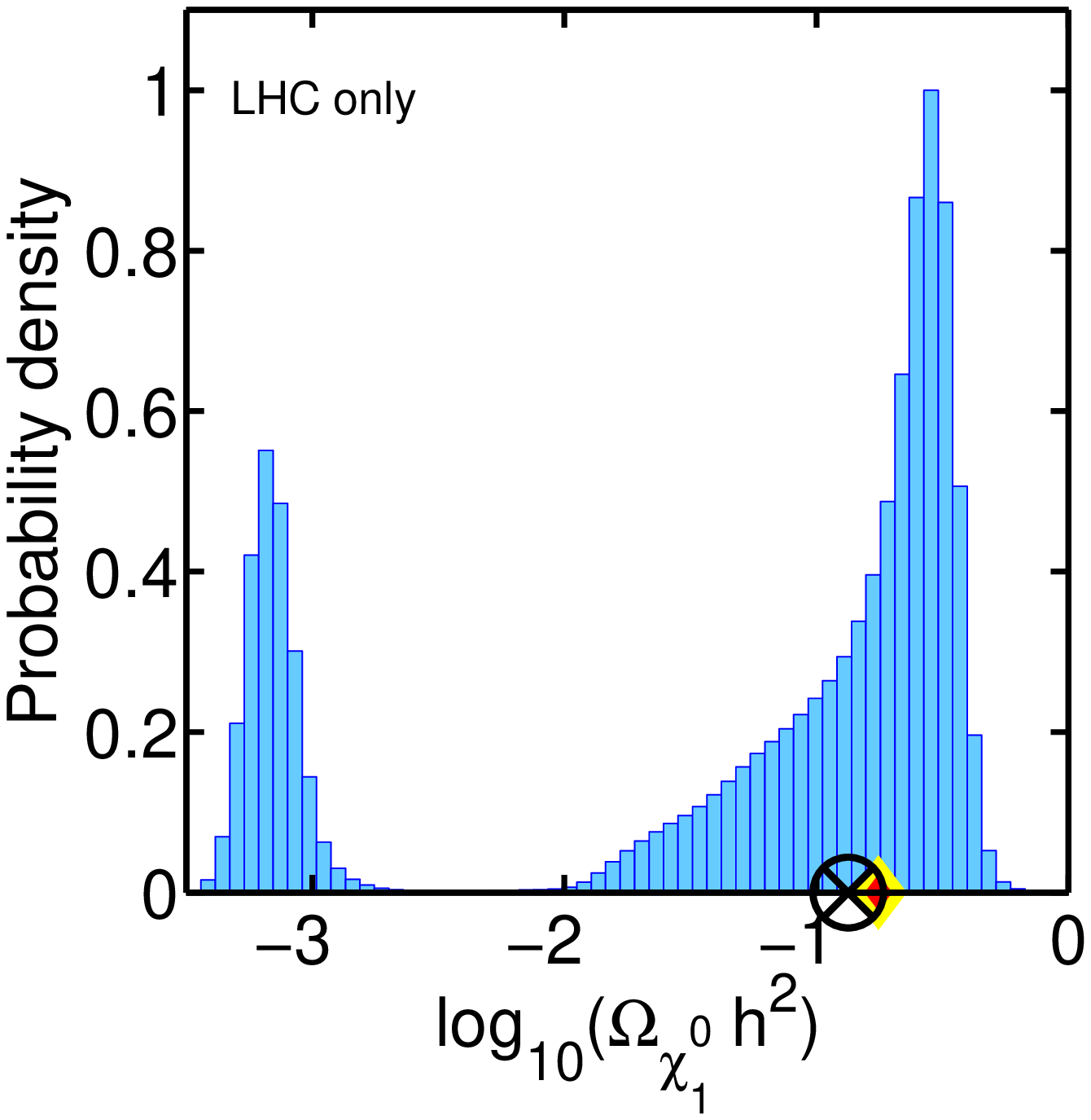,width=0.35\textwidth} 
\hspace*{-0.5cm}
\epsfig{file=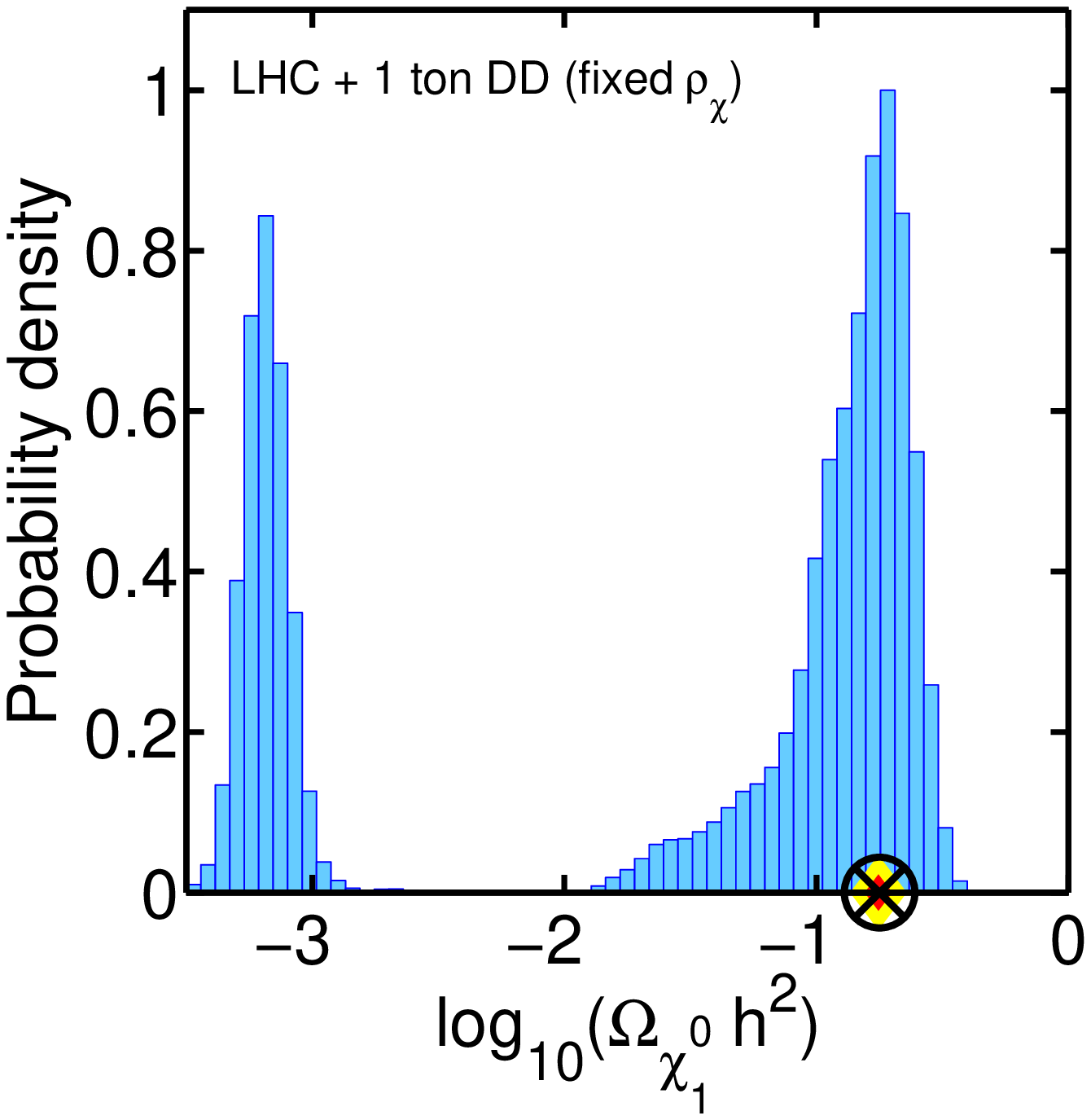,width=0.35\textwidth}
\hspace*{-0.5cm}
\epsfig{file=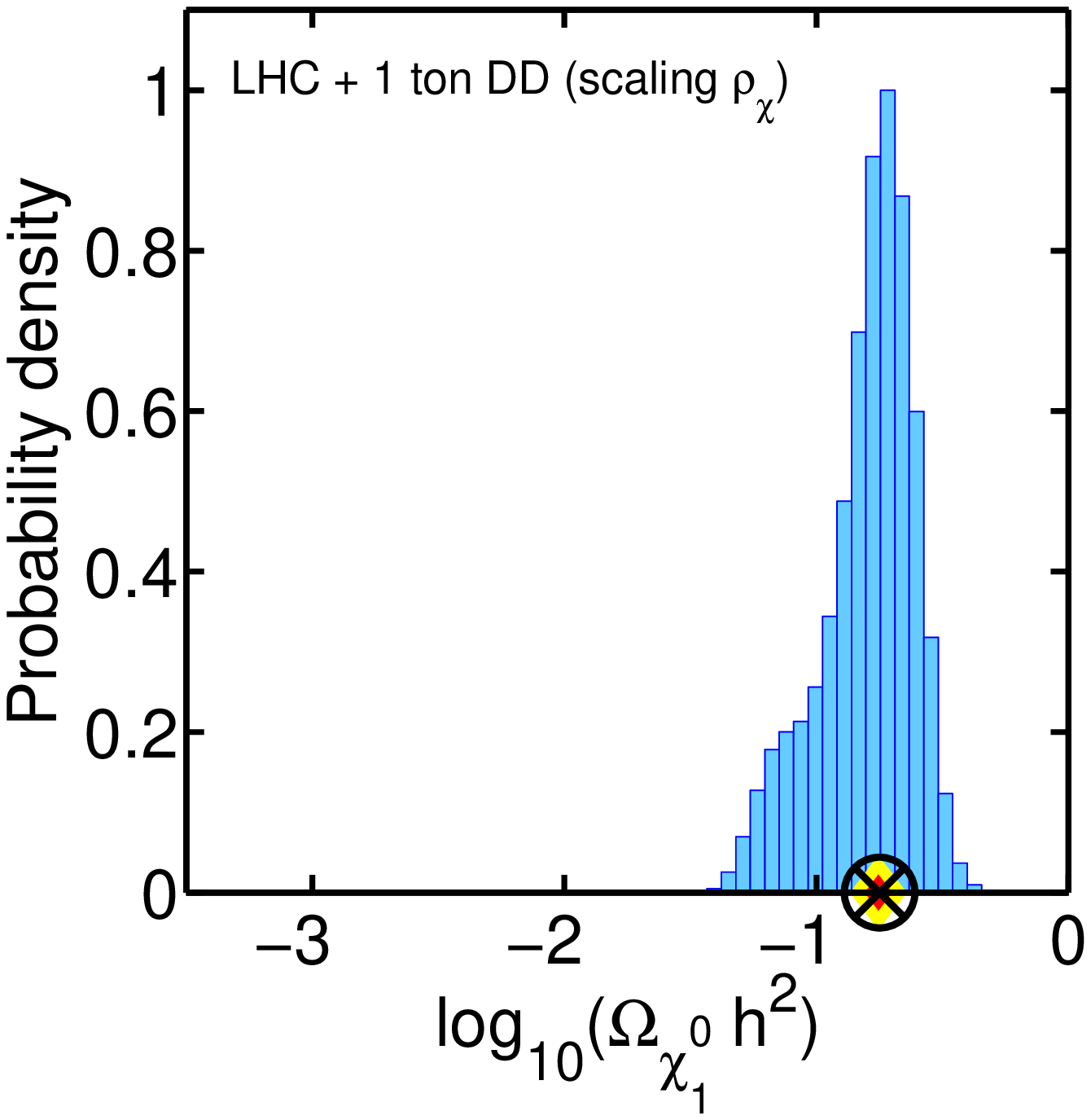,width=0.35\textwidth}
\vspace*{-0.7cm}
\caption{\label{fig:relic_density} Pdf of the neutralino relic density obtained from LHC data only (left panel), LHC plus direct detection data for a fixed local density (central panel) and LHC plus direct detection data under the {\it scaling Ansatz}. The best fit point is shown by the encircled black cross, while the true value is given by the yellow/red diamond.}  
\end{figure*}

\begin{figure*}[t]
\hspace*{-0.5cm}
\hspace*{-0.5cm}
\epsfig{file=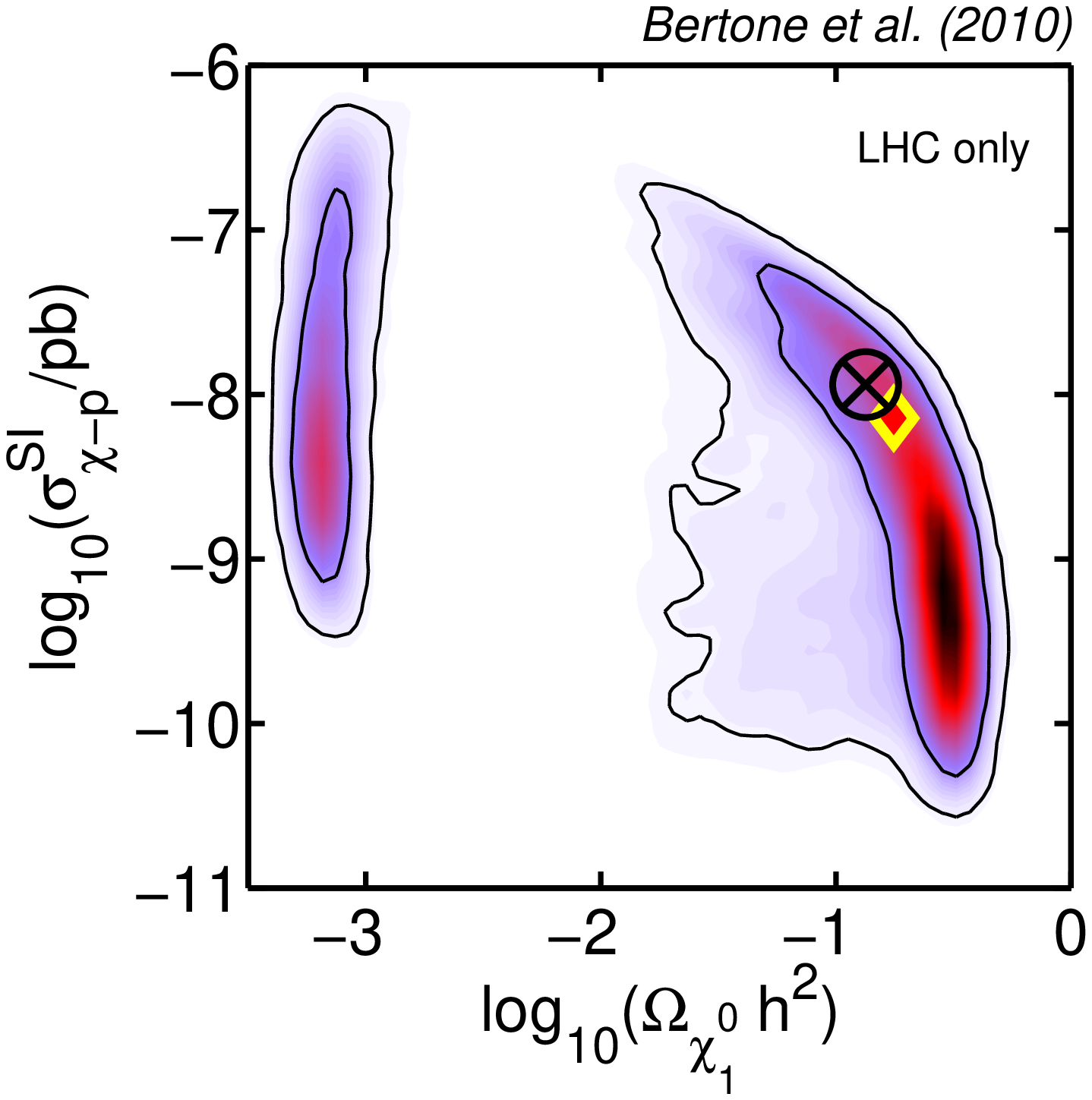,width=0.35\textwidth}
\hspace*{-0.5cm}
\epsfig{file=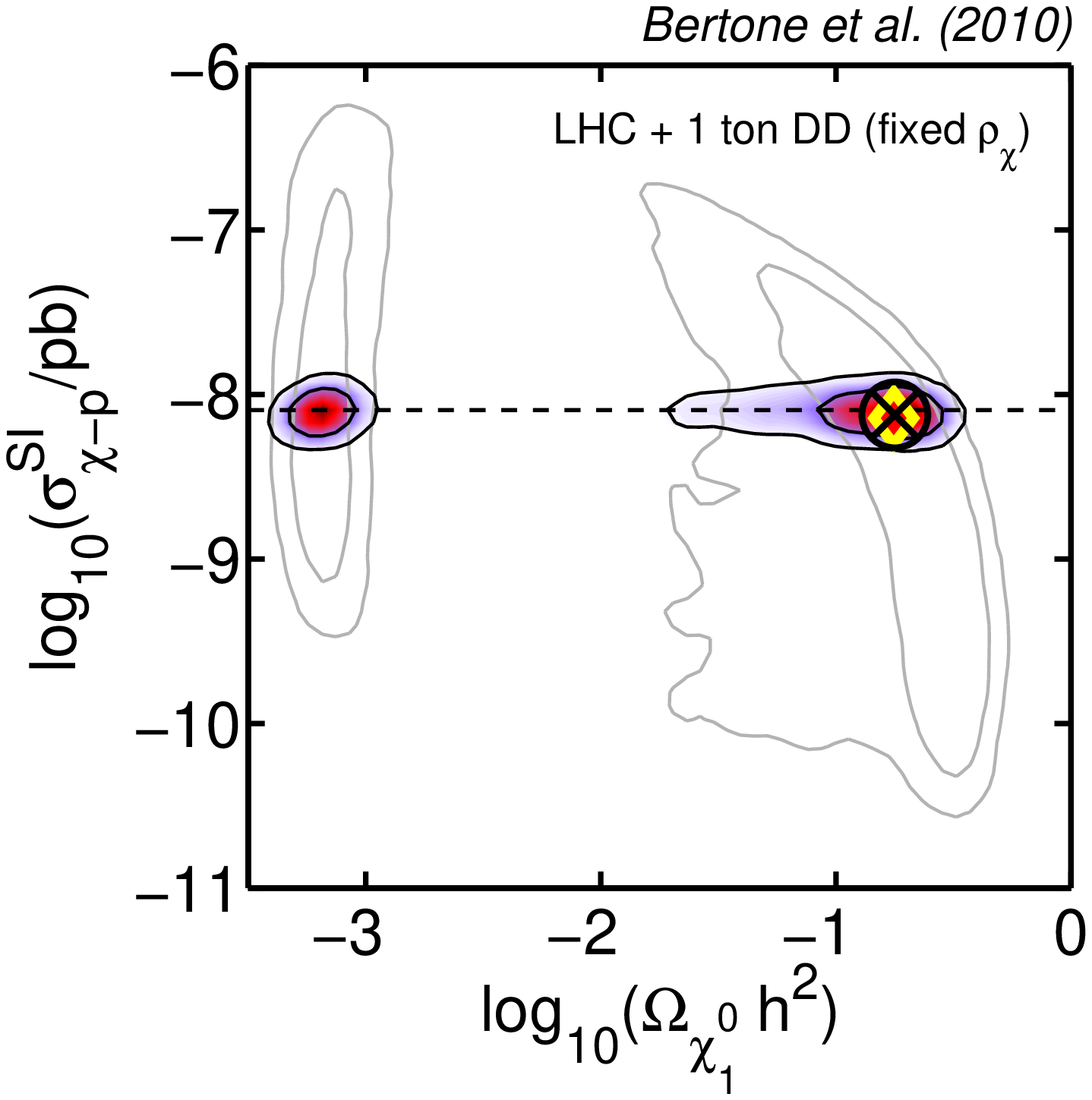,width=0.35\textwidth}
\hspace*{-0.5cm}
\epsfig{file=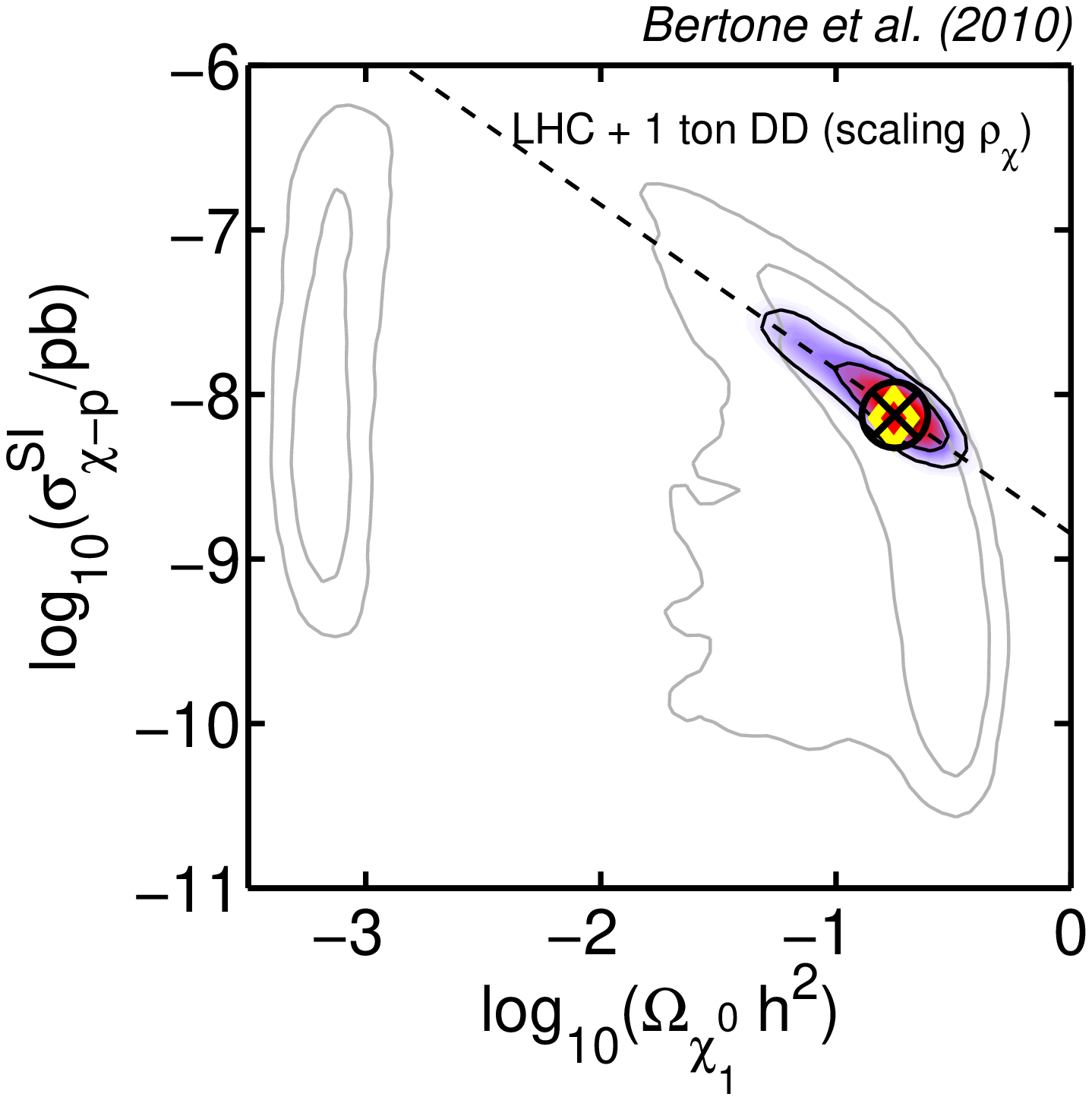,width=0.35\textwidth}
\vspace*{-0.7cm}
\caption{\label{fig:omega_vs_sigma} As in Fig.~\ref{fig:relic_density}, but in the $ \crosssection $ vs $ \relic h^2 $ plane. From left to right: constraints using LHC data alone, adding direct detection data assuming a fixed local DM density, and with our scaling Ansatz for the local density. The inner and outer contours enclose 68\% and 95\% probability regions, respectively. The probability distributions have been smoothed with a Gaussian kernel for display purposes. The best fit point is shown by the encircled black cross, while the true value is given by the yellow/red diamond. In the center and right panels we show for reference the LHC-only contours of the left panel (light grey), along with the directions along which the different Ans\"atze break the degeneracy (dashed lines). }
\end{figure*}

The reconstruction of the neutralino relic density is shown in 
Fig.\,\ref{fig:relic_density}. The left panel corresponds to the case where 
{\it only} LHC constraints are considered. Consistently with previous 
analyses \cite{Baltz:2006fm}, multiple peaks can be observed, as a 
consequence of degeneracies in the SUSY parameters space that the LHC 
constraints are unable to break. In particular, the two observed peaks 
correspond to neutralinos with different composition: mostly Wino and mostly 
Bino, from left to right. This is a consequence of the fact that the LHC 
is assumed to be able to measure only the two lightest neutralino states, 
but not the two more massive ones or the charginos. The true value of the relic density for our benchmark point ($\relic h^2 = 0.176$), represented by a diamond in Fig.~\ref{fig:relic_density}, is indeed inside the peak corresponding to mostly Bino dark matter. Although this value is about 60\% larger than the relic abundance measured by the WMAP satellite~\cite{Komatsu:2008hk}, we expect our results to remain qualitatively correct for other points in the co-annihilation region leading to the correct cosmological relic abundance. 
As commented above and already pointed out in previous works \cite{Baltz:2006fm} the better reconstruction of sparticle masses at the ILC could allow a more precise determination of the neutralino relic abundance, potentially removing some of these degeneracies. However, this information would only be available after a much longer period of time.

The constraints from LHC only data are also shown in the left panel of Fig.\,\ref{fig:omega_vs_sigma}, 
in the plane  $ \crosssection $ vs $ \relic h^2 $, where the true value of those quantities is given by ($\relic h^2 = 0.176$, $\crosssection = 7.1\times10^{-8}$ pb).
The leftmost region corresponds to a neutralino which has a leading
Wino component, thereby displaying a smaller relic abundance, whereas
the region towards larger relic abundance corresponds to Bino-like
neutralinos, for which the scattering cross section is also slightly
smaller. 

In the central and right panels of the two figures, we show the impact
of adding information from direct detection experiments. These plots have been
obtained by statistical posterior re-sampling of the LHC only scan, adding the
relevant Ans\"atze and the likelihood function of a direct detection experiment
as specified above. The central panels correspond to the assumption 1, or 
{\it consistency check}. This amounts to fixing the local 
neutralino density, and therefore 
we expect that only regions along a direction of 
constant $\crosssection$ to survive after direct detection 
data are implemented. 
This can be understood as follows: for a 
given number of measured events, and a fixed local density, there is only a range of values of $\crosssection$ that 
are compatible with the measurement. Notice that, as explained above, 
this Ansatz does not further constrain the neutralino thermal relic density. 
In this case the pdf for the neutralino relic density still 
displays the two maxima, corresponding to the two peaks in 
Fig.\,\ref{fig:relic_density} and the two ``islands'' in 
Fig.\,\ref{fig:omega_vs_sigma}. This is due to the fact that the neutralino 
can have a similar scattering cross section for both compositions and 
therefore (if the fact that it might be a subdominant DM component is not properly 
taken into account) could account for the same detected rate.

The most interesting case is the one that corresponds to our
assumption 2, namely the  
{\it scaling Ansatz}, which represents the most important result of this 
paper. When the appropriate scaling of the local density is applied, 
the Ansatz cuts the parameter space along a direction 
\begin{equation}
\crosssection \propto \relic^{-1},
\end{equation}
due to the fact that for a fixed number of events 
$ \crosssection \propto \rho_\chi^{-1} $ and that under the scaling Ansatz 
$ \rho_\chi \propto \relic $.
The dramatic consequences of this simple Ansatz are shown in the
right-most  
panels of both figures. Models corresponding to a low relic density are 
essentially ruled out, because under the scaling Ansatz they correspond to a low local density. Given a number of observed events in direct detection searches, a low local density would require a larger scattering cross section, which is incompatible with LHC constraints. As a consequence, the parameter space region corresponding to a neutralino that is mostly Wino can now be ruled out with high confidence, thereby leading to a much better reconstruction of the DM composition than it would be possible under the consistency check Ansatz.

We note that if the reconstructed relic density matches the observational determination of $\Omega_{\rm \tiny{DM}}$, this procedure also validates the standard cosmological history, and constrains deviations from the standard expansion rate at the epoch of DM freeze-out. Conversely, a mismatch between the reconstructed relic density and $\Omega_{\rm \tiny{DM}}$ would point towards a multi-component DM sector, or a non-standard expansion rate (see e.g. Ref.~ \cite{Gelmini:2008sh}  and references therein).

\section{Discussion and conclusions} 
We have investigated the effect of combining information from accelerator 
searches with data from direct DM detection, assuming realistic measurements 
at the LHC and in a Germanium detector with an exposure of 300 ton days. 

An interesting 
question is whether the systematic and statistical errors on this quantity and on other relevant 
physical quantities entering in the calculation of the event rate for direct
detection experiments can spoil the reconstruction procedure presented here. 
For instance, we 
have assumed a Maxwellian distribution for the velocity dispersion of DM 
particles, but a more refined analysis should keep into account the 
uncertainties on this quantity. Fortunately,  recent estimates based on numerical simulations, suggest that the small measured deviations from a Maxwell-Boltzmann distribution lead to errors of 10\% or less on the recoil rate, and they are therefore subdominant with respect to other uncertainties, such as the error on the nuclear form factor \cite{Ellis:2008hf}, and especially the error on the observed DM local density. The most important effect of such uncertainties, once marginalized over, would be to widen the pdf's of Fig.~\ref{fig:omega_vs_sigma} in the vertical direction by less than 10\%, if one considers only the statistical error on $\rho_{\rm \tiny{DM}}$ derived in Ref. \cite{Catena:2009mf}, and by up to a factor of two if one also considers the systematic error due to halo triaxiality \cite{Pato:2010yq}. Since the vertical thickness of the contours in the 2D posterior of Fig.\,\ref{fig:omega_vs_sigma} is approximately equal to a factor of 2, we expect that including these uncertainties would not modify qualitatively the marginal posterior distribution for $\relic h^2$, and our results would still apply. A more detailed discussion of these effects is 
beyond the scope of this paper, and we leave it to a separate upcoming work. However we explicitly studied the effect of varying the value for the mass of the top, including it as a nuisance parameter in the likelihood. The variation in the reconstructed pdf for the neutralino relic abundance and neutralino-nucleon scattering cross section is negligible.  

We stress once more the importance of combining different types of 
experiments.  The specific case discussed here shows that when reasonable 
assumptions are made to link the local density to the relic abundance, {\it a 
combined analysis of data from accelerators and direct detection experiments allows a
significantly better reconstruction of the DM properties}. 

This is true in 
the co-annihilation region discussed here, but it will provide important 
information for any SUSY scenario, and more in general for any new physics 
scenario. Even in cases where the LHC data are sufficient to pinpoint the 
underlying DM scenario, direct detection experiments can corroborate the 
results, and they can also be used to identify deviations from the standard 
expansion rate of the Universe at freeze-out that 
would appear as an inconsistency between the $\relic $ inferred from LHC 
data and cosmological measurements.

{\em Acknowledgements.} D.G.C. is supported by the Ram\'on y Cajal program of the Spanish MICINN, by the Spanish grants FPA2009-08958, HEPHACOS S2009/ESP-1473 and by the EU network PITN-GA-2009-237920.
R.T. would like to thank the EU FP6 Marie Curie Research and Training Network
``UniverseNet'' (MRTN-CT-2006-035863) for partial support, the Instituto de Fisica Teorica (Madrid) and the Institut d'Astrophysique de Paris for hospitality. The work of R. RdA has been supported in part by MEC (Spain)
under grant FPA2007-60323, by Generalitat Valenciana under grant
PROMETEO/2008/069 and by the Spanish Consolider Ingenio-2010 program
PAU (CSD2007-00060). We also thank the support of the spanish MICINN's Consolider-Ingenio 2010 Programme under grant MultiDark CSD2009-00064.

\end{document}